# Reform and Practice of Computer Application Technology Major Construction and Development in Higher Vocational Colleges in China -- Taking Jiangxi Vocational College of Applied Technology as An Example

*Yufei Xie, Yue Liu, Fan Zou*
*Jiangxi College of Applied Technology, Ganzhou 341000, Jiangxi, China*

Abstract: This study takes the development path of computer application technology specialty construction in Higher Vocational Colleges under the background of high-level higher vocational schools and specialty construction plan with Chinese characteristics (double high plan) as the main research object, and puts forward the core concept of computer application technology specialty construction and development in Higher Vocational Colleges in China through the practice of computer application technology specialty construction and development reform in recent years The main measures and construction objectives provide specific experience and solutions for deepening the reform of computer application technology specialty in higher vocational colleges.
Key words: China; Higher vocational colleges; Computer application technology major

1. INTRODUCTION
In recent years, with the rapid development of information technology at home and abroad, as well as the internal and external motivation of high-level higher vocational schools and specialty construction plan with Chinese characteristics (the "double high" plan), Jiangxi Vocational College of applied technology, on the basis of the comprehensive development of information specialty, has carried out multi angle, multi-level and multi-channel work, such as specialty construction, teaching reform, promoting learning by competition, and integration of production and education It has gradually become the information technology fulcrum for the construction of national "double high" construction specialty group, the demonstration platform for the "1 + X" reform of higher vocational information technology education, the training base for the preparation of information technology events of various skills competitions at all levels and Ganzhou Electronic Information Industry Co., Ltd It is the source of technical talents for the development of smart home industry and smart home industry.

2. THE CORE IDEA OF THE REFORM AND PRACTICE OF COMPUTER APPLICATION TECHNOLOGY
Through theoretical study and teaching practice, the teaching team of computer application technology in our school has formed a relatively systematic teaching thought and concept in the aspect of computer application technology, which is mainly manifested in the philosophical orientation of seeking truth from facts, the application orientation of learning for application, and the practical orientation of combining knowledge with practice. These three complementary and consistent It is also a kind of new idea.

2.1 Combine dialectical materialism with computer science teaching
In the course teaching, we should constantly strengthen the education of dialectical materialism philosophy. With the idea of seeking truth from facts, we should teach, compare and confirm the concepts of computational science, such as the relationship between computer hardware and software, the understanding of variables and functions, and dialectical materialism philosophy, such as the relationship between material and consciousness, the understanding of things and laws. In teaching, the collection and mining of information and data are summarized as the abstraction and refinement of social and economic development phenomena. On this basis, big data programming and calculation are the generalization of the regular understanding of the numerous and complicated social phenomena, unifying the learning process of computer big data technology with the historical materialism of "cognition practice re cognition re practice". In the teaching of computer course, students are fully aware of the scientific nature and correctness of dialectical materialism and historical materialism. At the same time, it also promotes students' comprehensive understanding and deep understanding of abstract computer theoretical concepts.

2.2 Always combine the new achievements of international computer teaching and research with the teaching and research reform of our university
In the first half of 2016, python, as the dominant language in the field of mathematical statistics, is still a new thing in Jiangxi Province, especially in the field of computer teaching in higher vocational colleges. On the basis of studying the frontier development of international computer teaching, our teaching team resolutely proposed the project "Research on the teaching reform of higher vocational computer programming language based on Python real-time simulation game system", which was officially approved in July 2016. From the perspective of wide application and strict standardization of Python language, this reform focuses on the analysis of the reform process of computer teaching represented by the United States, especially the transformation process of entry-level programming language from C language to Python language, and the application of code combat (https://github.com/codecombat/codecombat) And other open source computer real-time simulation game system for higher vocational students programming, In the process of programming language teaching in more than one teaching class in the following year, we have done a lot of in-depth and detailed comparative research, fully demonstrated the rationality, efficiency and applicability of the new teaching mode of Python language introduction combined with game system in computer programming language teaching, and published the relevant research results as the paper "reform of computer" Programming Language Teaching in Higher Vocational Colleges Based on the Characteristics of Python Language Grammar> Xie Yu-Fei, Zhong Dong- Bo. Reformation of Computer Programming Language Teaching in Higher Vocational Colleges Based on the Characteristics of Python Language Grammar[J]. Advances in Higher Education,2017, 1 (1): 26-30. which has laid a solid foundation for the follow-up teaching reform and the construction of big data technology and application specialty.

2.3 Combine the practical ability of information project development with the recognition of students' learning effect

Over the years, the teaching team's criteria for students' mastery of knowledge and skills have always been based on whether they can solve practical problems and meet the needs of users. It not only emphasizes the unity of theory and practice, but also emphasizes that practice is the only criterion for testing truth. In the scoring standards of teaching courses, project-based and work-based graduation design accounts for at least 50%. Every student who has passed the examination must be able to use the teaching programming language to develop and design a small program, solve a small problem, provide a small function, and see the big from the small, step by step, continuously improve their technical confidence and learning confidence, and constantly cultivate their talents Computer technology application level and ability.

3. MAIN MEASURES FOR THE REFORM AND PRACTICE OF COMPUTER APPLICATION TECHNOLOGY MAJOR

3.1 International standards for professional construction

As the world's fastest-growing field of science and technology, the rapid development of information technology also drives the continuous reform of information technology education. In the top ten universities in the United States, python programming language has become the most popular language for teaching introductory computer science courses. Among them, eight of the top ten universities' computer science departments use python, and 27 of the top 39 universities use Python [2]. As China's computer technology education has been following the development of education in Europe and the United States, there is a situation of secondary promotion. Until 2020, most of the entry-level computer language in higher vocational colleges are still using C language, a small number of them gradually start to use Java as the entry-level programming language, and only a few use pythons. Many higher vocational computer teachers are still very unfamiliar with Python. In order to build a high-level specialty with Chinese characteristics, we must learn, learn from and adopt the international cutting-edge teaching reform ideas and teaching methods, learn from each other's strong points to make up for the shortcomings and establish a new standard according to the actual development of computer education in China, and on the basis of integrating the cutting-edge achievements of computer education development, we can overtake and leap forward development In this case, we should explore a new path and mode to realize higher vocational education and even applied technology-based undergraduate computer professional education.

3.2 Teaching reform and implementation of "1 + X" project

In the national "double high" construction plan, "1 + X" certificate system reform is the leading hand in the direction of talent training and teaching reform, and it is also the basic, source and strategic reform of the national vocational education comprehensive deepening reform and lifelong vocational skills training system [3]. The formulation of the vocational skill level standard is not only a comprehensive clarification of the professional post responsibilities and vocational skill standards, but also a comprehensive normative guidance for the reform direction of vocational skill education and the talent training standards. In the next stage of teaching reform of computer application technology major in our university, we must fully implement the "1 + X" reform project in the field of computer application technology, take the vocational skill level standard in the field of computer application technology as the teaching standard, and take the five major subdivision categories launched by the Ministry of education in the field of computer application technology as the development direction. At the same time, we should focus on building a teaching reform and development mode with computer big data technology and application as the center, fully support the development direction of characteristic specialty groups such as natural resources big data and geographic information big data, and make more beneficial exploration and Practice for the Ministry of education to take the lead in proposing the vocational skill level standard with the characteristics of our school's professional development.

3.3 Build a professional team by promoting learning through competition

Taking the 19 disciplines competition of China's Higher Education approved by the world skills competition and the China higher education society as the scope of entry, we will take the national vocational college skills competition as the foundation, and take the innovation and entrepreneurship competition such as "Internet plus" and "Challenge Cup" as auxiliary, to explore the breakthrough of ACM-ICPC International Collegiate Programming Contest, and the information technology of occupation world competition (In). Formation and communication technology). We have gradually established a selection and cultivation system to adapt to the talent cultivation mode of higher vocational education. On the basis of ensuring the fairness of education and opportunities for all students, we have enabled students with ability, perseverance and strength to stand out, to obtain more comprehensive technical skills through more time and more efforts, and to prove themselves by participating in various competitions the third is the teaching path. In the process of professional construction and development, through all kinds of competitions at all levels, the young teachers who want to work and can work together to form a competition to promote learning, to promote teaching, to complement each other and to improve together Good competition mechanism.

3.4 Industry education integration serves local economy
The development of higher vocational education and the development of local industrial clusters are interdependent and help each other. Under the guidance of the concept of industry education integration and school enterprise cooperation, we need to enhance the initiative and service in exploring the mode of school in factory and factory in school, actively enter the industrial park and economic development Zone, investigate the development needs of local and enterprise, and constantly calibrate the talent training mode, so as to promote the professional construction In the process of deepening reform, we should constantly explore new models and new paths for deepening school enterprise cooperation. At the same time, we should improve the pattern, broaden our vision, actively connect with the "Dawan district" and other areas leading the reform and opening up, absorb the advanced experience of domestic and international school enterprise cooperation, and serve the national strategy of deepening the reform and development of vocational education.

4. CONCLUSION
After years of efforts, the computer application technology major of our university has basically realized the upgrading and reform of the talent cultivation concept and teaching methods of "cross-border integration, collaborative education, competency based, equal emphasis on morality and technology, individualized teaching, and integration of knowledge and practice", and has built the computer application technology major into an important supporting College for our university to undertake the task of national "double high" plan - land and resources investigation and management major group To build it professional and technical talents training highland to support the development of "Ganzhou electronic information industry belt", build a national ICT industry technology "double qualification" teacher training demonstration unit with relevant cooperative enterprises, focus on big data, cloud computing, Internet of things, artificial intelligence, mobile Internet and other new technology applications and "1 + X" certificate system, and jointly realize the integration of learning, training and teaching Finally, the education system and stable operation mechanism of school and enterprise, professional and industrial subject coordination, identity coordination, standard coordination, teacher coordination, curriculum coordination and environment coordination are constructed.